# BLOCKCHAIN APPLICATION IN SIMULATED ENVIRONMENT FOR CYBER-PHYSICAL SYSTEMS SECURITY

Riccardo Colelli    Chiara Foglietta [y]   Roberto Fusacchia [z]  Stefano Panzieri [x]  Federica Pascucci [{]

## ABSTRACT

Critical Infrastructures (CIs) such as power grid, water and gas distribution are controlled by Industrial Control Systems (ICS). Sensors and actuators of a physical plant are managed by the ICS. Data and commands transmitted over the network from the Programmable Logic Controllers (PLCs) are saved and parsed within the Historian. Generally, this architecture guarantees to check for any process anomalies that may occur due to component failures and cyber attacks. The other use of this data allows activities such as forensic analysis. To secure the network is also crucial to protect the communication between devices. A cyber attack on the log devices could jeopardize any forensic analysis be it for maintenance, or discovering an attack trail. In this paper is proposed a strategy to secure plant operational data recorded in the Historian and data exchange in the network. An integrity checking mechanism, in combination with blockchain, is used to ensure data integrity. Data redundancy is achieved by applying an efficient replication mechanism and enables data recovery after an attack.

## 1  Introduction

Industrial control systems (ICS) are linked Cyber-Physical Systems (CPS) used for the management and monitoring of critical activities, through the use of sensors and actuators, controlled by programmable logic controllers (PLCs) and Supervisory Control and Data Acquisition (SCADA) systems. A Historian is a device, present in industrial control systems, that has the purpose of receiving, analyzing and saving the data and commands transmitted on the network, through the PLCs, to detect any process anomalies that may occur due to the failure of devices. For this purpose, data stored in the Historians will be used as input for offline analysis activities, such as forensic analyzes.

Attacks against Critical Infrastructures (CIs) weaken the functioning of a country and they have a serious impact on the safety of the population. In the last decades, many SCADA systems are increasing their communication capabilities, so that they have greater flexibility and ease of management. The use of network communication brings inherent vulnerabilities that can be exploited by an individual, or group actors to carry out actions that undermine the proper functioning of the industrial control system and consequently of all the infrastructure monitored by it. Recent incidents, such as Stuxnet [1], the power blackout in Ukraine [8] and Trisis [11], have highlighted these dangers and increased the importance of cybersecurity in these systems.

In this paper, a defensive strategy for ICS in CIs is proposed and analyzed. This strategy can be adopted in the industrial field to protect data and processes from increasingly frequent cyber attacks. The technique in question is based on and is inspired by the now-famous blockchain technology, introduced by Satoshi Nakamoto in 2008 [17], as a solution to

    Riccardo Colelli is with Department of Engineering, University of Roma Tre, 00146 Rome, Italy, riccardo.colelli@uniroma3.it

    [y]Chiara Foglietta is with Department of Engineering, University of Roma Tre, 00146 Rome, Italy, chiara.foglietta@uniroma3.it

    [z]Roberto Fusacchia is with Department of Engineering, University of Roma Tre, 00146 Rome, Italy, rob.fusacchia@stud.uniroma3.it

    [x]Stefano Panzieri is with Department of Engineering, University of Roma Tre, 00146 Rome, Italy, stefano.panzieri@uniroma3.it

    [{]Federica Pascucci is with Department of Engineering, University of Roma Tre, 00146 Rome, Italy, federica.pascucci@uniroma3.it

the possibility of carrying out online transactions without the need for third-party intermediaries to guarantee their correctness, in particular, to ensure the transmission of digital currency.

For a time associated exclusively with cryptocurrency, in recent years the potential of blockchain has been fully perceived and it has been understood that the possible uses of this technology are not just limited to finance. The blockchain is a data structure whose entries are grouped into blocks, concatenated in chronological order. The feature that makes this innovation special is the immutability, which it possesses thanks to an articulated system that guarantees extreme redundancy of these blocks, whose integrity is ensured by the use of cryptography.

The immutability of blockchain is suitable for solving many current problems, including the manipulation, by malicious actors, of operational data. The unauthorized modification in the industrial field in automated processes aims of creating harmful malfunctions. Therefore, a new possible architecture is presented, designed to improve data security over three dimensions: immutability, confidentiality and redundancy. This methodology is extended to the communication between devices to avoid Data Injection. Finally, data redundancy is obtained by applying an efficient replication mechanism that allows data recovery after an attack.

## 1.1 Paper Contribution

The methodology used in this paper aims at giving a blockchain architecture for CPS to secure communication between devices. The contributions of the paper are two-fold. First, we apply the blockchain architecture in a simulated environment. A simulated environment reduces the time and costs deriving from a real system during the test phase. Second, the proposed tool is tested under a cyber-attack. In particular, we analyze the data manipulation in the Historian database and the Data Injection through a Man-In-The-Middle attack. In this paper, preliminary results are presented in the framework of Cyber-Physical security for critical infrastructure scenarios.

## 1.2 Organization of the Paper

The paper is organized as follow. In Section 2 related works and contribution are considered. In Section 3 a blockchain architecture integrated into a SCADA system with Mininet simulation environment is presented. In Section 4 the basic case study, two water tanks system, is described. Results are presented in Section 5, where manipulation in the database and injection between two hosts are considered. Conclusions and future works are drawn in Section 6.

## 2 Related Works

In literature, different approaches have been proposed for blockchain in Cyber-Physical Systems.

Our architecture is motivated by [15], where authors propose an architecture to secure plant operational data recorded in the Historian. Process historians are being used to store data originating from a variety of sources in SCADA, including control and monitoring, laboratory information management and asset management systems. For this reason, the role of a Historian is crucial and an integrity checking mechanism, in combination with blockchain, is used to ensure data integrity.

In [6], authors propose a tamper-free plant operation system by applying blockchain technology to the integrated plant management system in a nuclear plant. The purpose of applying blockchain technology is to manage the registration and processing results of sensitive information objectively and transparently.

Authors in [9] implement blockchain in the data acquisition part of SCADA systems in the area of the smart grid with a personalized mining node selection process. Blockchain is also applied in a smart grid scenario in [13]. The use of blockchain in IoT s points of contact with the physical world has already been demonstrated in [7], where authors use peer-to-peer systems and smart contracts. Proper of the industrial IoT is the architecture proposed in [19], where authors describe the key technologies, for the blockchain-based smart factory.

In [16], blockchain technology applications are used to securing the smart grid in a distributed energy context. Also, in [12] a distributed data protection framework based on blockchain is proposed for securing power grid. The use of blockchain could be vertical inside industry 4.0: authors in [14] analyze the product lifecycle management dedicated to integrate the information inside the enterprise and realize the information and service sharing cross-enterprise. In [5], authors propose a comparative analysis for abnormal events in industrial processes and find the architecture that provides the best anomaly detection capability.

## 3   Proposed method

The blockchain is a possible solution to the platforms that solve our needs in an innovative way or those of the companies or public administrations that provide the services we use. In particular, the blockchain refers to themes and concepts of digital innovation, which are: trust, responsibility, community, decentralization.

In blockchain technology, each node is connected to all other nodes and there are no central servers or gateways. The main elements that make up the blockchain architecture are nodes, transactions, blocks, ledger and hash. Nodes are the participants in the blockchain and are physically constituted by the servers of each participant. Then, the transaction consists of the data representing the values subject to exchange and which need to be verified, approved and then archived. Block is represented by the grouping of a set of transactions that are combined to be verified, approved and then filed by the participants in the blockchain. Finally, the ledger is the public register in which all transactions carried out in an orderly and sequential manner are noted with maximum transparency and in an immutable way. The ledger consists of the set of blocks that are chained together through an encryption function and thanks to the use of hashes. Hash is the operation that allows to map a text and/or numeric string of variable length into a unique and univocal string of determined length. The Hash uniquely and securely identifies each block and must not allow tracing back to the text that generated it.

Each block of the chain can contain a certain number of transactions, which concern the exchange of digital assets, and use a peer-to-peer network that stores these transactions in a distributed manner [18]. The actors who own digital assets and transactions involving a change of ownership are registered within the block through the use of public and private key cryptography and digital signatures that guarantee security and authenticity to the exchange. Each block has an identifying hash value so that it is uniquely and securely recognized: it is structured in such a way as to prevent the reconstruction of the text from which it was generated. In addition, each block having its identifying hash also contains the hash of the block that precedes it, so that when a new block is added to the chain, the blockchain can maintain a shared and agreed view of the current state. One contribution of this paper is the integration between blockchain and Mininet for the simulation of CPS networks. With the Mininet it is possible to simulate multiple nodes on a network and virtually connect them with switches and links. Every node simulates a stand-alone machine with its network features. Mininet program [4] was used, more specifically an extension of it: MiniCPS [2]. These two programs allow you to simulate the communication between devices using the Modbus industrial protocol, to have a scenario that is as realistic and relevant to reality as possible.

## 4   Case study

To demonstrate the advantages of using the blockchain and encryption in the Operational Technology (OT) environment, we decide to simulate an industrial system for the control of a physical process to ensure the integrity, availability and confidentiality of data. This process consists of maintaining a predetermined water level in two tanks. This scenario has been implemented on a virtual machine in Linux operating system. In the simulation, the traffic of TCP Modbus packets was also emulated between different devices in the industrial network. For data storage and database management, including simulation of attacks against them, sqlite3 [10] was used.

Finally, to test the response of the network to the False Data Injection attack, Scapy [3] was used. Scapy is a packet manipulation tool for computer networks, capable of capturing, falsifying or decoding packets and sending them over the network. Furthermore, Scapy is also able to handle tasks such as scanning, tracerouting, unit tests, indeed, network attacks.

### 4.1   Network structure

The industrial network implemented in this paper is structured as follows. At the field level there are two tanks, each of which is equipped with a sensor for measuring the water level (S1 and S2); three valves, one at the inlet to tank1 (A1); one between the two tanks (A2) and the last outlet to tank2 (A3), can be opened or closed to regulate the water flow between the tanks. The opening and closing of these actuators are completely automated and managed by two PLCs, which, based on the measurements received from the sensors, decide what to do. As shown in Fig. 1, the PLC1 receives the measurements from sensor1 (S1), therefore respective to the water level present in tank1, and controls valve A1, while PLC2 receives the measurements relating to tank2, measured with sensor2 (S2), and manages valves A2 and A3. In a legacy network for automation, the values collected by the PLCs, in addition to being processed locally for the management of the valves, would be sent one at a time to the SCADA system. These values would be saved as well as they are in the Historians and shown to the operator through the HMI (Human Machine Interface).

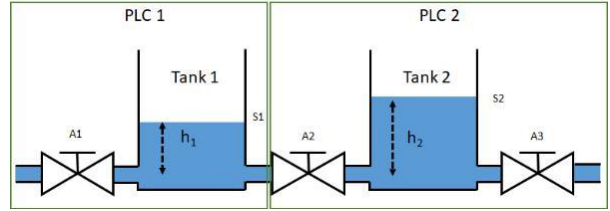

Figure 1: The physical process of two water tanks controlled by the two PLCs.

As opposed to the traditional way of storing data, in this paper, a double signature system was introduced for the encryption and authentication of messages exchanged between the devices. Furthermore, storage nodes were added to the system to having a database within them that will function from Historian. Hence, these nodes will store the data received from the PLCs and they have the task of creating and managing replicas of this data. In this way it will be possible to store a single data in several Historians, creating a fundamental redundancy for its availability in forensic analysis or for backup actions. Finally, to better protect this data, a private blockchain has been implemented within the network, in which the only node capable of generating a new block is the blockchain module. The condition that makes this happen, is that at least one of the messages received by this module are authentic. Thus, messages were sent by a storage node. The number of storage nodes inserted in the simulated network is equal to 6 and it has been chosen to replicate the data with a factor of 3 (i.e. each measurement to be stored will be replicated in 3 Historian). Furthermore, both the storage nodes and the blockchain module, due to the double signature system, can detect and report to operators any tampering and verify the authenticity of the messages. A summary of the structure implemented in this project is shown in Fig. 2. Storage nodes are composed of three modules:

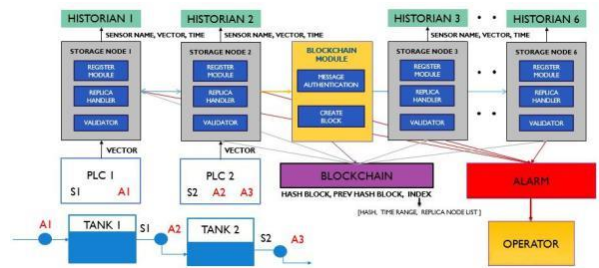

Figure 2: The functioning of the network with blockchain.

- Register module: has the task of storing the measurements detected by the sensors, with the corresponding time range, in the Historian.
- Replication handler: replicates the vector stored in the Historian, sending it to two other pre-selected storage nodes.
- Validator: verifies the integrity of all data stored in the Historian using the blockchain. If the validator finds a manipulation of the values stored in the Historian, it will notify the operator and start the automated data recovery process. The following paragraph will analyze in detail the network presented here, to better explain the functioning of the various systems present in it.

4.2 Functioning of the network

The operation of the structure in the paper is therefore as follows:

Step 1: The sensors collect the data relating to the water level present in the tanks.

Step 2: The PLC processes these measures to decide if and which valves to open.

Step 3: To avoid wasting memory, having to memorize on each device the entire blockchain with its content, which provides a hash index for each encoded measurement. The PLC does not send every single value to the storage node, but rather inserts it into a vector, which will contain all the measurements captured by the sensor in the predetermined time interval, in this case one minute.

Step 4: At the end of the time interval, the PLC encrypts the created vector using the public key of the recipient node storage (so that only those in possession of the recipient's private key, hopefully only the recipient himself, can decrypt it) and, subsequently, generate the fingerprint through the hash function; then it also encrypts the resulting digest, this time, however, using its private key (to create a digital signature that guarantees the authenticity of the sender). Finally, it sends both encrypted messages to the storage node through the TCP Modbus.

Step 5: The storage node that receives the two messages decrypts the first using its private key and the second using the sender public key. Then it generates in turn the vector fingerprint received by the PLC and compares it with the one obtained by decoding the second message. If the two digests do not match, it means that either a device error or an attempted sabotage by a malicious agent has corrupted the data and it will therefore make no sense to store it. In fact, not only would they be useless, but they would also become misleading in case of subsequent analysis. The storage node will then discard the message and warn the operator in the control room that something is not working. On the other hand, if the network is functioning correctly and no one has manipulated the message, the two hash values will correspond, guaranteeing that nothing has been altered and that these measures are those detected by the sensor placed on the tank.

Step 6: Having checked the integrity of the data received, the storage node generates a second hash function, encoding inside it, in addition to the vector of measurement received, the name of the sensor that detected them and the date with the time in which they were taken, to ensure this vector a unique fingerprint for the entire existence of the system . Then it will encrypt and sign this digest, as explained above for the PLC, to send it in a secure way to the blockchain module.

Step 7:The blockchain module will verify the authenticity of all messages sent to it within the predetermined time interval by the various storage nodes. Then, if there is at least one valid one, it will undertake to create a new block in the blockchain, in which it will insert the following values:

- Indexes: it will store in each block as many indexes as they have been authenticated during the one minute interval. Within each index there are: i) The hash string that represents the fingerprint of the vector that has been stored in the Historian; ii) The date and time when the measurements within the vector were taken; iii) The three identifiers of the Historians in which the vector will be stored. The first value in this list indicates the database in which it has already been registered, while the other two, selected randomly, determine those in which the replica will have to be inserted,

- Hash identifier of the block: which uniquely identifies the block just created, and which is calculated starting from the hash of the last block stored in the blockchain,

- Reference to the hash of the previous block: identified of the last block stored in the blockchain. Thanks to this value it will be possible to trace, one by one, all the blocks present in it.

Moreover, the blockchain will be immutable, both because it will be stored inside each device present in the network, and because the blockchain module is the only device authorized to insert blocks into the and, as we have seen, it will only be able to do so if the message is authentic (double signature verification). In this way it is sure that these blocks are correct and can be used later in the data integrity check phase in the Historians.

Step 8: Once the blockchain module has inserted the new block into the blockchain, it sends a log message to all storage nodes, inside which the digest that identifies the last block inserted in the blockchain will be encoded with the respective public keys of each of them.

Step 9: Having received this log, each storage node decrypts the message through its private key and uses it to query the blockchain, thus obtaining the indices present in the last block of it. From these it will extract the list of Historians selected for storing the replica of the vector, the second and third indexes present in the list, and check if one of these corresponds to its own. If that is the case, it will have to request the vector to be stored inside, from the node that has already registered it previously, which the one whose index is at the top of the list. This exchange of data also takes place in complete safety and confidentiality by exploiting asymmetric cryptography in the way already extensively discussed above, once obtained the vector will store it in its Historian.

Step 10: Once all the vectors have been replicated where necessary, each storage node will check cyclically, if the data stored in its database is still available and intact. To do this, each of them will scroll through all the blocks stored in the blockchain, to access all the digest stored in it. Therefore, for each transaction recorded in a single block, the storage node accesses the index that contains the hash string, the date of storage and the list of nodes in which the vector was stored and, if in this list it recognizes its Historian, it checks all Inside it, on the date indicated, a vector is stored whose fingerprint is identical to that stored in the blockchain index. If this check gives an affirmative result for all the blocks in the chain, each measurement stored up to this moment in the Historian is available and intact.

Step 11: On the other hand, if the hashes do not match, the automated instant recovery system is activated, which consists in requesting the replication of the damaged data to another storage node, among those indicated in the replica nodes, to back up and restore the vector in memory.

These phases are replicated in the same order with maximum synchronization to ensure that everything is running smoothly. Therefore, the difference between a standard industrial network and the blockchain industrial network described is the insertion of storage nodes and the use of the blockchain to keep track of all the data stored in the Historians. The tools used to implement and simulate this structure in a virtual environment and the attacks with which it was tested will be described in more detail below.

## 5   Results

To assess the resilience of the network, the strategies that a malicious actor could use to compromise it will be analyzed and implemented. The outcomes resulting from two types of cyber attacks will be presented: data manipulation in Historians and Man-In-The-Middle (MITM) attack between two devices of the network with subsequent false data injection.

### 5.1   Data manipulation in the Historian

In this first scenario, the attacker will modify the data already stored in the database, which could later be used to carry out forensic analyzes. In particular, data detected by the sensor1 and stored in a precise instant within the Historian1 will be modified, as shown in Fig. 3(a). We are assuming that the attacker managed to obtain the authorization to be able to modify the records in the database. In this case, the effect of the aforementioned attack can be observed, for instance, in the vector associated with the measurements detected by the sensor1 and stored in the Historian1, subsequently modified by a different vector through a database upload. However, in the implemented network, an automated data recovery system is provided. In particular, once a minute, the validators present in each storage node will compare the digests stored in the blockchain with those recreated by interrogating the Historian and, if the two do not match, the guarantee of the immutability of the blockchain allows the storage nodes to realize that tampering has occurred within your database.

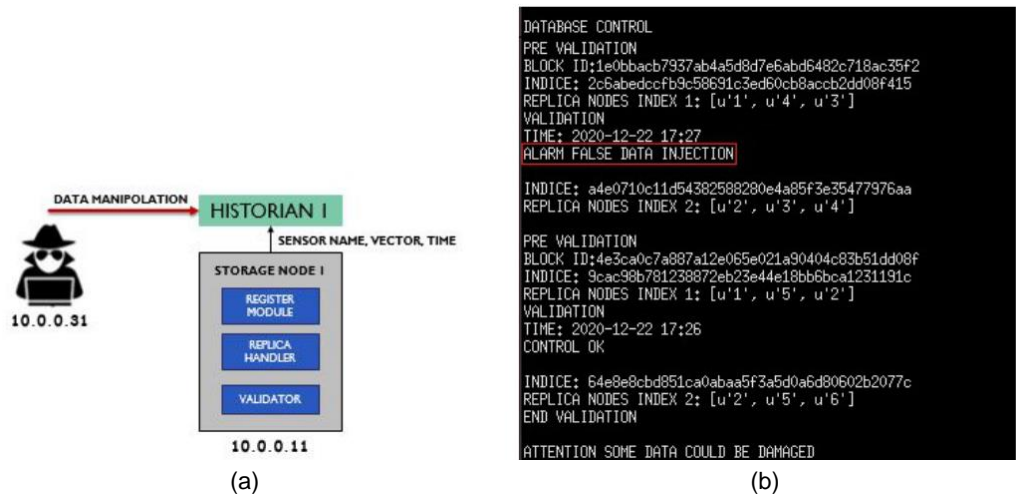

Figure 3: Data manipulation attack against Historian (a) and manipulation detected (b).

An example of the effect that such an attack can have on the network is shown in Figure 3(b), which shows the behaviour of the storage node1 after changing the vector related to the measurements detected by the sensor1 stored in the Historian1. We assume that the attacker can access and modify values from Historian. In particular, he/she manipulates value as shown in Table 1 in order to obtain the values reported in Table 2.

The node first performs a pre-validator call in which, by querying the blockchain through the hash block previously received with the log, it manages to obtain the indexes stored within the block identified by that string. As can be seen from Figure 3(b), in the first index obtained there is, within the replication nodes, the value 1, which identifies the Historian1. Since this index is relative to the measurements taken at 17:28, it indicates that in the Historian1, there must be a vector collected by the sensor1 at that precise time (i.e. index 1 in the first place).

Table 1: Data stored inside the Historian1.

| Name | Value | Time |
|------|-------|------|
| Sensor 1 | [2,5] | 17:26 |
| Sensor 1 | [6,7,7,6,7,7,6,7,7,6] | 17:27 |
| Sensor 2 | [4,4,5,4,5,3,6,3,6,3] | 17:28 |

Table 2: Data manipulation inside the Historian1 without blockchain countermeasures

| Name | Value | Time |
|------|-------|------|
| Sensor 1 | [2,5] | 17:26 |
| Sensor 1 | [2,1] | 17:27 |
| Sensor 2 | [4,4,5,4,5,3,6,3,6,3] | 17:28 |

At this point the node queries its database to check the integrity of that vector and, once the array has been obtained, negates the fingerprint, i.e. the digest which must be identical to that stored in the blockchain; if not, the data within Historian1 has been manipulated. Thus, it is observed that this check has given positive feedback ('check ok').

Subsequently, the validator also checks the second index present within the block of the blockchain considered, but since the value 1 is not present in the list of replication nodes, it did not have to perform any action. Check, therefore, all the indexes present in the first block analyzed, through the 'prev_block' field present in each block of the blockchain, the storage node1 goes back to the previously stored block and obtains the saved indexes, to carry out the aforementioned operations again.

This time, however, it is analyzing the record hit by the attack: the validation detects that the data within that record have been falsified ('False Data Injection alarm') and automatically activates the recovery system. In this regard, the storage node1 will ask the storage node6 (second index in the 'replica nodes' list) to send the vector that the latter should have stored in its Historian. The message will always be sent through asymmetric cryptography to guarantee its integrity and confidentiality.

Once received the carrier, the validator will reconstruct the digest again using the hash function and if it coincides with the one stored in the blockchain, it will use the carrier received to replace the damaged one inside of its Historian. If, however, once again the two indexes were to differ, the storage node1 will still have an opportunity to retrieve the data by requesting the vector from storage node3, the third ID in the 'replica nodes' list.

Once the vector has been restored, the validator will continue to trace the entire blockchain, through the 'prev_block 'field, until it has checked all the blocks belonging to it, performing the same operations for each block. The storage node will signal to the operator in the control room through a message, if the process went well or if it was necessary to carry out the recovery. In this second case, it will be the operator's task to investigate the causes that led to the damage to the data detected.

5.2  False Data Injection

The second attack tested is in the Man-In-The-Middle scenario, in which it is assumed that the attacker can interpose between two devices of the network. A wired Man-In-The-Middle (MITM) attack in MiniCPS is launched by a malicious device. Data Modification starts when the adversary changes in the TCP packets only the bytes of the payload related to the sensor readings, as shown in Figure 4.

Firstly, in the communication between the PLC1 and the storage node1. Secondly, the attacker is in between the storage node1 and the blockchain module. In this way, it will be able to capture and store all the data packets exchanged between these devices, then retransmit them as they are in order not to arouse suspicion.

We assume that an attacker has compromised communication between two nodes (e.g. PLC1 and storage node). The adversary has limited knowledge of our system, i.e. he/she knows the physical model we use, but he/she doesn't know the thresholds we select to raise alerts. Moreover, the attacker has obtained adequate knowledge through eavesdropping, becoming able to perform the cyber-attacks described in the following. Given this knowledge, he/she generates a data injection attack with the goal of sensor and actuator data tampering. The detection statistic will always remain included in the space of the selected threshold. We assume that the adversary is undetected by the network security systems (e.g. Intrusion Detection System), in particular, that the attacker has already gained access to the control network.

We assume that the attacker is not able to modify all measures consistently, he/she can only want to change a subset of the system measures.

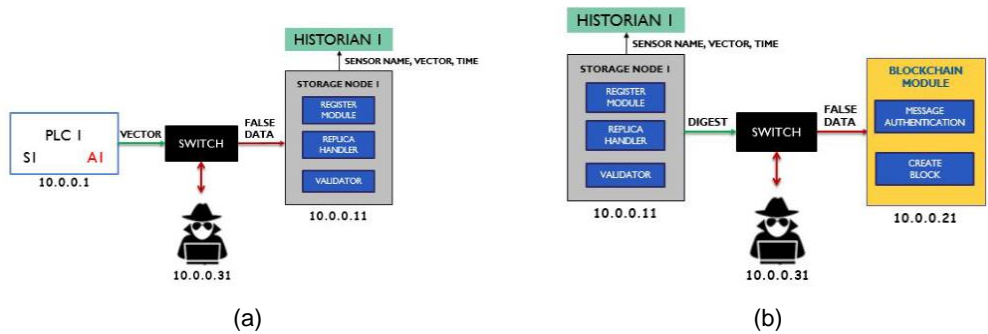

(a)                         (b)

Figure 4: False Data Injection attack between PLC1 and storage node1 (a) and between storage node1 and blockchain module (b).

(a)                         (b)

Figure 5: Detection of False Data Injection attack between PLC1 and storage node1 (a) and between storage node1 and blockchain module (b).

Considering that, the message is encrypted, the attacker cannot understand the data content. However, this does not discourage him/her and to create damage to the attacked system, he/she wants to inject false measures into the network to compromise the forensic analyzes or to create false alarms in an attempt to slow down or block the process. Therefore, the double signature system implemented in the network will allow the receiver of the message to recognize the tampering by comparing the hash code calculated from the manipulated message arrived at its destination and the one that arrived via a message encoded with the private key of the sender. The attacker who manipulates the message is not aware of the sender's private key. Consequently, the device does not recognize the received data packet as authentic and generates an alarm promptly reporting the anomaly to the operator, as shown in Figure 5(a).

The same type of attack is replicated, between the storage node1 and the blockchain module, with very similar results, as shown in Figure 5(b). However, the blockchain module does not store it within the blockchain because the digest received from storage node1 is not authentic. The issue can be promptly reported to the operator. Therefore, in the absence of the index within the blockchain relating to the stored vector, the validator will not be able to check its integrity which consequently, for this reason, cannot be guaranteed.

# 6 Conclusion

Our work finds possible allies in blockchain technology and asymmetric cryptography to guarantee the security of these systems from cyber attacks. In particular, in this work, it was demonstrated how it is possible to implement an architecture resilient to attacks, such as Man-In-The-Middle with False Data Injection. We implement the network for the automation process by using Mininet and MiniCPS simulators. It can be seen from the results that the proposed network architecture based on the blockchain can better mitigate these attacks.

Furthermore, the proposed architecture guarantees the availability of data and integrity and confidentiality, thus, allow the continuity of the process and the possibility of carrying out a posteriori analysis to make the system more efficient and solve the problems encountered.

However, the role of operators remains crucial, as they monitor all the processes occurring in the network in the control rooms. Hence, even if the effects of the attacks are mitigated by the implemented architecture, a prolonged attack can still lead to the loss of a lot of information and an overview of everything that is happening and the operator will be the only agent able to bring the situation back to normal, possibly in the shortest possible time, also foreseeing the effects that the attack may have had on the entire system and acting accordingly.

In the future, we can think of implementing blockchain, not only to defend the data present in the Historians but also for other devices present in a network larger than the one analyzed. Furthermore, a way can be found to solve the problem of retransmission of damaged data, as in the case analyzed, in which the attacker manages to interfere in the communication between two devices, affecting it through the damage of the data packets exchanged between the entities. An idea could be to establish a second redundant communication channel, which can be used if the first is compromised.

## 7 Acknowledgments

The current work has in parts been supported by the EU projects RESISTO (Grant No. 786409) on cyber-physical security of telecommunication critical infrastructure.